\documentclass{article}
\usepackage{epsfig}

\begin{document}

\title{Dephasing of entangled qubits} 
\author{Peter Nalbach\\ 
Department of Physics, Stanford University, Stanford, CA 94305, USA}
\date{\today}
\maketitle
\begin{abstract}
We investigate the influence of nearby two-level systems on
the dynamics of a qubit. The intrinsic decoherence is given by a
coupling of both the qubit and the two-level systems to a heat
bath. Assuming weak interactions between the qubit and the two-level
systems we show that the dephasing of the qubit follows a multi-exponential
decay. For a flat distribution of energy splittings of the two-level
systems the multi-exponential behavior simplifies to a good
approximation to a single exponential
decay. The rate is given by the sum of the dephasing rate of the
isolated qubit plus all relaxation rates for the two-level systems out
of their present state. This latter contribution sums up to a rate
approximately given by the mean coupling between the qubit and the two-level
systems times the number of thermally active neighbors.
\end{abstract}


In recent years, extensive studies have been done on quantum
computing. One of the major problems to overcome in that field is the
decoherence of the qubit dynamics. Besides the dephasing due to
experimental imperfections the interactions with the environment
yields an intrinsic source of decoherence
\cite{Ref1,Ref2,Ref3,Ref4}. The so-called {\sl quality factor} of the
qubit, the number of quantum operations performed during the qubit
coherence time, should be at least $10^4$ to allow for quantum error
correction\cite{Ref5}. We will not discuss the various physical
sources of decoherence. Instead we investigate a model where the qubit is
weakly coupled to defects described as two-level systems (TLSs). Since
the qubits are also described as TLSs, we do formally not distinguish
between qubits and defects. All TLSs/qubits are coupled to a heat bath
represented by a set of harmonic oscillators providing the basic source of
relaxation and decoherence. Both result from the
energy exchange with the heat bath like in the spin-boson model
\cite{Leggett} which is different from the model proposed by Unruh
\cite{Ref1}. Our investigation focuses on the
question how neighboring TLSs influence the dephasing behavior of a
given qubit. In the
following we use {\it qubit} to address the TLS of interest and {\it
TLS} for all the systems which form a background for the qubit. Qubits
which are not used for an actual calculation contributes in that
picture as a background.


We restrict our investigation to longitudinally coupled TLSs leading to
the Hamiltonian 
\begin{equation}\label{HS} H_S \,=\,-\sum_i \frac{\Delta_{i}}{2}\sigma_x^{(i)} 
\,-\sum_{i\not= j} J_{ij}\sigma_x^{(i)}\sigma_x^{(j)} \; 
\end{equation}
with the Pauli matrices $\sigma_i$, the energy splittings $\Delta_i$
and the interaction constants $J_{ij}$. We allow for distributed
splittings and couplings. The many-body eigenstates are
given by the product states of the isolated TLSs. Nevertheless, the
{\sl longitudinal} coupling changes the energies of the excitations,
thus leading to entanglement between the quantum states of the 
noninteracting qubits. 

The bosons of the heat bath,
$H_B=\sum_k\omega_kb_k^\dagger b_k$,
couples with their strain field linear to the {\sl dipole} operator
$\sigma_z^{(i)}$ of the two-level systems 
\begin{equation}\label{HB} H_{SB} \,=\, \sum_i\sigma_z^{(i)}\cdot\sum_k\,
\lambda^{(i)}_k(b_k+b_{-k}^\dagger) \; . 
\end{equation}
This coupling reflects the natural picture that the two-level
system can change its state via an absorption or an emission of a
boson. 

For a single qubit this is the widely investigated {\sl spin-boson}
model\cite{Leggett,Weiss}. The case of a transversely  coupled pair is
investigated by Terzidis et al. \cite{Orestis} and Dub\'e et
al. \cite{Stamp}. 
It is well know that a coupling to bosons leads to an additional
transversal coupling between the two-level systems of the form
$\widetilde{J}_{ij}\sigma_z^{(i)}\sigma_z^{(j)}$ through virtual phonon exchange
\cite{Jofrin,Kassner}.
We assume in the following that we can neglect that additional
coupling. 
We are mainly concerned in the limit of weak coupling between the
qubits and the heat bath. We focus on the dissipative dynamics
generated by the coupling to the environement and thereby, on the
dephasing times.
The renormalization of the excitation energies are beyond the scope of
the present paper. We treat the coupling within a resumed
perturbative approach \cite{Nal01a,Wuerger} as introduced by W\"urger
which is a weak coupling approach. 

For sufficiently dilute two-level systems, the time evolution of the
boson operators is not affected by the two-level systems and the
effect of the heat bath is entirely characterized by the coupled
density of states
\begin{eqnarray}\label{spectrum} J_i(\omega) &=& \frac{\pi}{2}\sum_k\,
|\lambda_k^{(i)}|^2\, \delta(\omega-\omega_k) \\
&\simeq& \pi\alpha^{(i)}\omega_c^{2-s}\omega^s \Theta(\omega_C-\omega) \; . \nonumber
\end{eqnarray}
with the cut-off frequency $\omega_c$ of the bath and a dimensionless
coupling constant $\alpha^{(i)}$. The right hand side of eq.(\ref{spectrum})
assumes a linear dispersion of the bosons with an sharp upper
cut-off. An exponent $s=1$ describes
systems exhibiting ohmic dissipation \cite{Leggett} and accordingly
the bath is called ohmic. An exponent $s>1$ is
usually called a super-ohmic bath where especially $s=3$ is relevant
for a defect in a solid interacting with acoustic phonons. 


We will start with a discussion of a pair of TLSs. Its spectrum
as plotted in figure 1 consists of four levels. Next to the
levels we give the states as product states of the ground (down) and
the excitated (up) state of the two TLSs. The full arrows give the
transitions induced by fields coupling to the dipole of the first TLS
and the dashed arrows the ones resulting from the second TLSs.

Let's assume that TLS 1 is our qubit and we prepare it in
a superposition between up and down. Accordingly TLS 2 is a defect.
For an isolated qubit the decay of
coherence is given by the probability of a boson absorption or
emission. At temperature below the energy splitting $\Delta_1$ of the
qubit, phonons with energy $\Delta_1$ are rare and the main effect is
given by spontaneous emission.
However, for a pair of coupled TLSs it is necessary to have
information on the state of the second TLS. If it
is in its down-state, boson absorption of the second qubit would also lead
to decoherence for the first qubit since the original (coupled) state
is altered.
If the second qubit is in its up-state, boson
emission of the second qubit contributes to the decoherence of the
first one. This reflects the entanglement between the two TLSs. For two
entangled qubits performing a quantum calculation, it
is obvious that the calculation is lost if one of the TLSs
dephases. Accordingly the dephasing rate of the pair is the sum of the
rates for the two isolated qubits. (This argument holds only true
as long as the interaction does not change the quantum mechanical
evolution of the qubits meaning $J\ll\Delta_i$.) Since the above
argument was independent of the strength of the interaction
between the TLSs, the question arises, how we can avoid entanglement of
a qubit with surrounding TLSs which do not take part in the actual
computation. 

To answer this question we derive the memory kernel, resulting from
the coupling to the heat bath, for the dynamics
of the dipole operator $\sigma_z^{(i)}$ of the first qubit. Following the
approach in ref. \cite{Nal01a} we calculate the memory kernel within
the lowest contributing order of the coupling between system and bath
and apply a Markov approximation. Neglecting the real part of the
memory kernel, which represents the renormalization of the energy
splittings, we end up with the rates for the various dynamic
processes. 
We restrict ourselves to the subspace of the operators,
$|\hspace*{-1mm}\uparrow\downarrow\rangle\langle\downarrow\downarrow\hspace*{-1mm}|$ and
$|\hspace*{-1mm}\uparrow\uparrow\rangle\langle\downarrow\uparrow\hspace*{-1mm}|$, which correspond
to a transition from down to up for the first qubit or, in other
words, to a preparation in a superposition between up- and down-state.
The dynamics of this subspace is independent from the
rest space and the time evolution of the subspace of
the complex conjugate operators is identical. 
 
The dynamics of the operators
$|\hspace*{-1mm}\uparrow\downarrow\rangle\langle\downarrow\downarrow\hspace*{-1mm}|$ and
$|\hspace*{-1mm}\uparrow\uparrow\rangle\langle\downarrow\uparrow\hspace*{-1mm}|$ are not
independent from each other since the off-diagonal components of the
memory kernel are non zero. 

The time-evolution operator in Laplace space for this subspace (in
the basis of the above given operators) is given
by
\begin{eqnarray} && {\cal U}(z) \,=\,\\ && \left(\begin{array}{cc}
z - \omega_{+} + i(\Gamma_1+2\Gamma_{2\uparrow}) &
-i2\Gamma_{2\downarrow} \\ 
-i2\Gamma_{2\uparrow} & z - \omega_{-} +
i(\Gamma_1+2\Gamma_{2\downarrow}) 
\end{array} \right)^{-1} \nonumber 
\end{eqnarray}
with the excitation energies $\omega_{\mp}=\Delta_1\mp
2J\approx\Delta_1$, the usual one-phonon rate of the first qubit
$\Gamma_{1}=J_1(\Delta_1)\coth(\beta\Delta_1/2)$, the decay rate
of the second qubit due to phonon emission
$\Gamma_{2\downarrow}=J_2(\Delta_2)(1+n(\Delta_2))$ with the Bose factor
$n(\Delta_2)$ and the decay rate of the second qubit due to phonon
absorption $\Gamma_{2\uparrow}=J_2(\Delta_2)n(\Delta_2)$. Thereby
$J_{1/2}(\Delta_{1/2})$ are the coupled density of states for the two
TLSs (compare eq.(\ref{spectrum})). The phase coherence rates are
given by the imaginary parts of the two poles of the time evolution
operator. 

This dynamics has two simple extremes. When the
difference of the two excitation energies
$|\omega_{+}-\omega_{-}|=4J$ is bigger than the off-diagonal
components, we can neglect the off-diagonal entries. In this case, the
dynamics of qubit 1 has two different rates which depend on the state
of qubit 2. Therefore, the rates are determined by both processes:
decay of the qubit 1 and decay of the qubit 2.

In the second case,
$J<\mbox{max}\{\Gamma_{2\downarrow},\Gamma_{2\uparrow}\}
=\Gamma_{2\downarrow}$, we have to taken
into account the off-diagonal entries. The off-diagonal
elements cause that the eigenstates of the time evolution do not
correspond to the
eigenstates of the Hamiltonian. Instead the dynamics of the two qubits
completely decouples and the coupling is negligible (for the dynamics
of the qubit 1) as long as it is smaller 
than the linewidth of the states of the isolated qubit 2 which is
given by $(\Gamma_{2\uparrow}+\Gamma_{2\downarrow})/2$.

Thus, only when the coupling between a
qubit and a second TLS is smaller than the linewidth of the second TLS
the dynamics of the qubit is independent from the second TLS. We
should point out that the actual quantum mechanical evolution of the
states are only weakly affected since $J/\Delta\ll 1$. But the
dephasing rate might be changed considerably since we have no
knowledge  about $\Gamma_1\hspace*{0.0em} ^>\hspace*{-0.65em}
_<\Gamma_{2\downarrow/\uparrow}$.


In the case of many qubits the above conclusion remains qualitatively
valid. The decoherence rate of a qubit of
interest depends on the actual states of all other TLSs. If we have a
distribution of interaction energies, we have
to devide all TLSs into two groups depending on the
condition $J_{\alpha i}<\Gamma_{i\downarrow}$ or $J_{\alpha
i}>\Gamma_{i\downarrow}$ where $\alpha$ is the index of the qubit of
interest.

If $J_{\alpha i}<\Gamma_{i\downarrow}$ the i-th TLS does not influence
the qubit $\alpha$. Nevertheless a flip 
of the qubit $i$ will change the resonance energy of the qubit
$\alpha$. Since that change is smaller than the linewidth, we would
not be able to measure it. In the case of many neighbors
fulfiling that condition, the changes might add up resulting in
a change in the resonance energy which is bigger than the
linewidth. In an echo experiment at an ensemble of qubits, this leads
to an 'effective' dephasing where the relative phase
between the qubits decays. This effect is called 'spectral diffusion'
\cite{KlauAnd68}.

All neighbors in the second group with $J_{\alpha
i}>\Gamma_{i\downarrow}$ form an entangled cluster with the qubit
$\alpha$. As in the case of a pair, the decoherence rate of a
superposition of eigenstates of the qubit $\alpha$, with all the TLSs
in an eigenstate, gets:
\begin{equation} \Gamma_t \,=\, \Gamma_{\alpha} \,+ 2\hspace*{-4mm}\sum_{i \in \atop \{i|J_{\alpha
i}>\Gamma_{i\downarrow}\} }\hspace*{-4mm} \Big\{ \begin{array}{cc} \Gamma_{i\uparrow}
& \mbox{TLS $i$ in the down-state} \\ \Gamma_{i\downarrow}
& \mbox{for TLS $i$ in the up-state} \end{array}
\end{equation}
Thereby, we assumed a clear separation between $J_{i\alpha}$ and
$\Gamma_{i\downarrow}$. If both are of the same order of magnitude, the
time evolution is a complicated mixture of both extremes. This
assumption should at least be valid for the majority of the TLSs.
If the TLSs are qubits like the qubit $\alpha$, the decoherence
rate scales with the number of qubits. This seems 
to be natural for a situation where $N$ entangled qubits perform one
calculation. If only one of these qubits dephases, the calculation is
lost. The main point lies in the fact that they are entangled as long
as the interaction between two qubits exceeds their linewidth. 
For example, if one wants to send quantum information from one to
another quantum computer, the interaction must be bigger than the
linewidth since otherwise, the quantum computers would dephase even
before they are entangled (starting from an isolated initial
condition). After the transmission both computers must be separated
again fulfilling the above condition.

Within the picture of one qubit surrounded by TLSs, we have to
specify the distributions of couplings $P(J)$ and of the resonance
energies $Q(\Delta)$ in order to get quantitative results.

Assuming homogeneously distributed TLSs and a dipole-dipole like
interaction between TLSs we obtain for the distribution of couplings 
\begin{equation} P(J) \,=\, \frac{\bar{J}}{J^2}
\end{equation}
between a minimal and a maximal value of the coupling with equal
probability of both signs. $\bar{J}\simeq NJ_{min}$ normalizes the
distribution to the number of TLSs $N$. Since
$\bar{J}\approx\langle|J|\rangle=\bar{J}\ln(J_{max}/J_{min})$, we refer
to $\bar{J}$ in the following as the mean coupling.

We want to consider a case in which all neighbors are
equivalent
\begin{equation} Q(\Delta) \,=\, \delta(\Delta-\bar{\Delta}) \; .
\end{equation}
Without further knowledge over the neighbors we expect them in thermal
equilibrium with the environment. In that case the average decay of
the qubit $\alpha$ follows the form ($J_i\ll\bar{\Delta}$)
\begin{equation} \langle e^{-\Gamma t}\rangle \,\simeq\, e^{-\Gamma_{\alpha} t}\cdot
\left\{ p^- e^{-2\bar{J} t} \,+ p^+ e^{-2e^{-\beta\bar{\Delta}}\bar{J}
t} \right\} 
\end{equation}
with the occupation probabilities of the ground/excited state
$p^{\pm}=\exp(\pm\beta\bar{\Delta}/2)/\cosh(\beta\bar{\Delta}/2)$.

This simple result has few remarkable features. First of all, we have
no longer a simple exponential decay but a sum of two. The original
decay of the qubit $\alpha$ is still present in the prefactor. The
decay of the TLSs lead to the term in the brackets. This term
does not depend on the actual decay rates of
the TLSs but only on their mean coupling $\bar{J}$ to the
qubit $\alpha$ and the probabilities of these neighbors to be in their
ground or excited state. If the temperature is higher than the energy
splitting 
of the TLSs ($k_BT\gg \bar{\Delta}$) both rates are equal to a good
approximation. In the other limit ($k_BT\ll \bar{\Delta}$) only one
decay is relevant and its contribution from the neighboring TLSs is
vanishing. Thus only thermally active TLSs are affecting the dephasing
of the qubit. Actually two decays are only seen in the case where
the temperature matches the energy splitting.

In the second step, you might consider neighbors with broadly
distributed energy splittings. If we assume, for example, just a flat
distribution $Q(\Delta)=Q_0 \Theta(\Delta_{min}-\Delta)
\Theta(\Delta-\Delta_{max})$ between a minimal and a maximal value, we
get 
\begin{equation} \langle e^{-\Gamma t}\rangle \,\simeq\, e^{-\Gamma_{\alpha}
t}\cdot \prod_i \left\{ p^-_i e^{-2\bar{J}_i t} \,+ p^+_i
e^{-2e^{-\beta\bar{\Delta}_i}\bar{J}_i t} \right\} 
\end{equation}
where $i$ runs over all neighbors and $\bar{J}_i=Q_0J_{min}$ with the
normalization factor $Q_0$. In order
to get simple analytical expressions we approximate the exponential by
$\exp(-\beta\bar{\Delta}_i)\simeq\Theta(T-\Delta_i)$. This
approximation fails for all systems with $\Delta_i\simeq T$. If the
majority of TLSs, as in the case of a flat distribution of energy
splittings, does not fulfill that condition, we expect that the
approximation works pretty well and we get
\begin{equation} \langle e^{-\Gamma t}\rangle \,\simeq\, e^{-\Gamma_{\alpha}
t \,-2\bar{J}(T) t} 
\end{equation}
with $\bar{J}(T)=Q_0(k_BT-\Delta_{min})NJ_{min}$.
This shows an additional dephasing rate which is approximately linear in
temperature as long as the temperature is between the extremal values
of the distribution $Q(\Delta)$. The dephasing due to nearby
local defects is given by 
the mean coupling to the thermal active neighbors. In fact, the above
used approximation divides the TLSs sharply in thermal active and
inactive ones. Thereby we loose the multi-exponential behavior of the
correct solution.


We investigated the dephasing in a model of weakly coupled TLSs. The
coupling was chosen only to influence the energies of the excitations
and thus to ensure entanglement. It does not change the states in a
sense that the eigenstates of a coupled cluster are the product states
of the isolated TLSs. In order to cause dephasing we coupled all TLSs
to a heat bath in a way that dephasing as well as relaxation is caused
through energy exchange with the bath. 

We find that the dephasing of a qubit weakly coupled to $N$ TLSs is no
longer a simple exponential decay but a complicated multi-exponential
decay. We investigated two simple cases. In both, we
assumed a homogeneous distribution of TLSs and a dipole type coupling
between the qubit and the TLSs. 

In the case of identical TLSs, we get two exponential decays
instead of one. If the temperature is higher than the energy splitting
of the TLSs ($k_BT\gg \bar{\Delta}$), both rates are equal. In the
other limit ($k_BT\ll \bar{\Delta}$) only one 
decay is relevant and its contribution from the neighboring TLSs is
vanishing. Thus only thermally active TLSs are affecting the dephasing
of the qubit. Actually the two decays are only seen in the case where
the temperature matches the energy splitting. 

On the other hand, a flat distribution of energy splittings for
the TLSs leads to a multi exponential decay. Nevertheless, we end up
with one major dephasing rate where all 
thermally active TLSs are contributing to. The dephasing through local
defects is thereby given by the mean interaction with all thermally
active defects. Thus, the additional dephasing is independent from the
actual decay mechanisme of the TLSs. 

The above results are valid as long as the linewidth of the TLSs is
smaller then their coupling to the qubit. This fact opens another way
to avoid dephasing due to nearby TLSs by increasing the linewidth of
the thermal active TLSs. 

The author wants to thank H. Horner, B. Thimmel, Y. Lee and
W. Harrison for many
stimulating and clarifying discussion. Parts of the work were done in
the Institut f\"ur Theoretische Physik at the University of
Heidelberg. I want to thank the DFG which supported that part within
the project HO 766/5-3 'Wechselwirkende 
Tunnelsysteme in Gl\"asern und Kristallen bei tiefen Temperaturen'. I
also want to thank the Alexander-von-Humboldt foundation which
supports the work done at Stanford.

\newpage

\epsfxsize=6cm
\epsffile{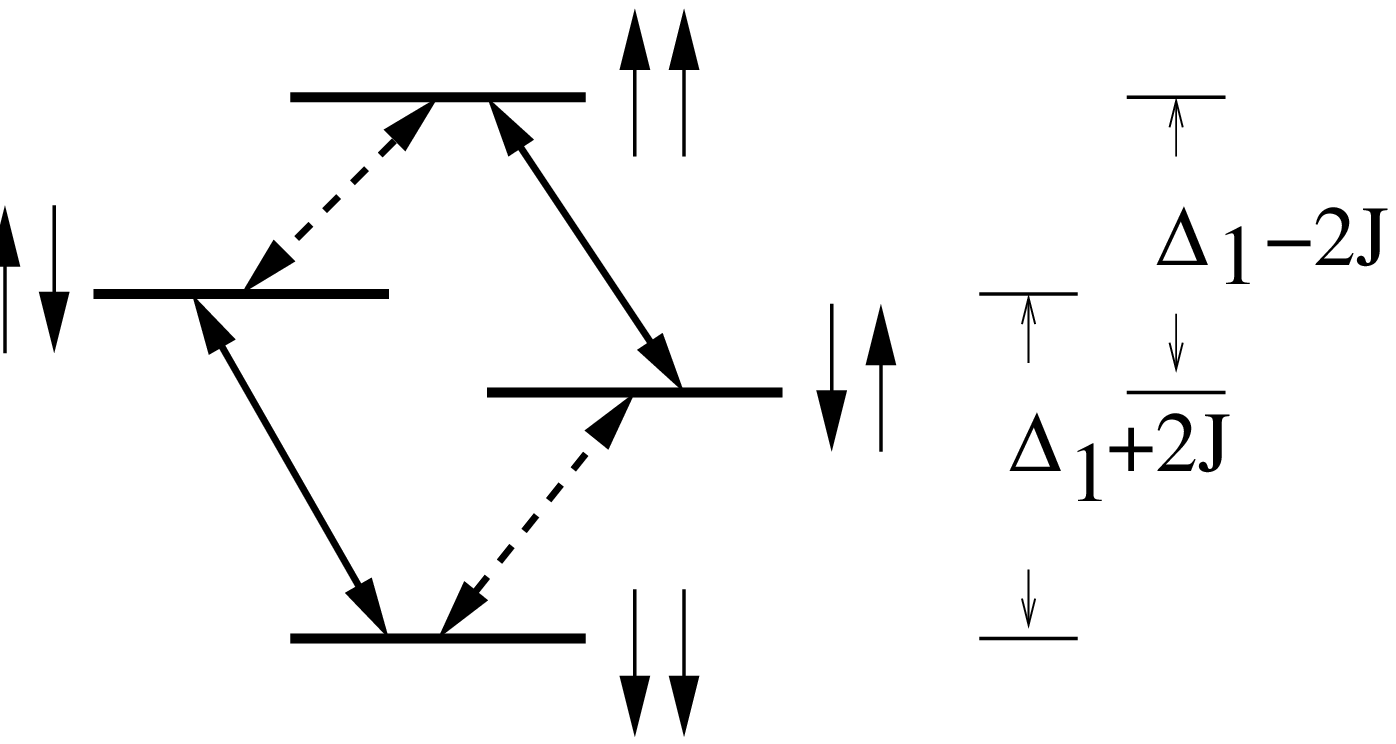}

Fig. 1: The spectrum of two coupled TLSs. The states are
labeled by their composition of the up and down states of the two
isolated TLSs. The possible phonon transitions are shown by the
arrows. The full (dashed) arrows are transitions induced by
interactions with the TLS 1 (TLS 2).


\begin{thebibliography}{99}

\bibitem{Ref1} W.G. Unruh: Phys. Rev. A {\bf 51} (1995) 992

\bibitem{Ref2} A. Garg: Phys. Rev. Lett. {\bf 77} (1996) 964

\bibitem{Ref3} G. Massimo Palma, K.-A. Suominen, A. Ekert:
Proc. R. Soc. Lond. A {\bf 452} (1996) 567

\bibitem{Ref4} Yu. Makhlin, G. Sch\"on, A. Shnirman:
Rev. Mod. Phys. (2001) in press

\bibitem{Ref5} J. Preskill: Proc. R. Soc. Lond. A {\bf 454} (1998) 385

\bibitem{Leggett} A.J. Leggett, S. Chakravarty, A.T. Dorsey,
M. Fisher, A. Garg, W. Zwerger: Rev. Mod. Phys. {\bf 59} (1987) 1

\bibitem{Weiss} U. Weiss: {\sl Quantum Dissipative Systems}, Series in
Modern Condensed Matter Physics, Vol. 2, World Scientific, Singapore
(1993) 

\bibitem{Orestis} O. Terzidis, A. W\"urger: J. Phys.: Condens. Matter
{\bf 8} (1996) 7303

\bibitem{Stamp} M. Dub\'e, P.C.E. Stamp: Int. J. Mod. Phys. B {\bf 12}
(1998) 1191;

\bibitem{Nal01a} P. Nalbach: to be submitted to Phys. Rev. B

\bibitem{Wuerger} A. W\"urger: J. Phys.: Condens. Matter {\bf 9}
(1997) 5543

\bibitem{Jofrin} J. Joffrin, A. Levelut: J. Phys. (Paris) {\bf 36}
(1975) 811

\bibitem{Kassner} K. Kassner: Z. Phys. B {\bf 81} (1990) 245

\bibitem{KlauAnd68} J.R. Klauder, P.W. Anderson: Phys. Rev. {\bf 125}
(1962) 912

\end{thebibliography}
\end{document}